\documentclass[12pt]{iopart}
\usepackage{iopams}
\usepackage{graphicx}

\newtheorem{rem}{Remark}
\newtheorem{rem1}[rem]{Remark}

\def\b{\beta}

\def\P{{\cal P}}

\newcommand{\teta}{\rlap{\lower2ex\hbox{$\,\tilde{}$}}\eta{}}

\def\P{{\cal P}}
\def\S{{\cal S}}

\def\ppt{t^{\prime \prime}}

\def\lp{{\ell}_{\rm Pl}}

\newcommand{\rcr}{\rho_{\mathrm{crit}}}

\newcommand{\f}{\frac}

\usepackage{colordvi}

\def\f{\frac}

\def\epsilon{\varepsilon}

\newcommand\be{\begin{equation}}
\newcommand\ee{\end{equation}}
\newcommand\ba{\begin{eqnarray}}
\newcommand\ea{\end{eqnarray}}

\begin{document}

\title{Are loop quantum cosmos never singular?}

\author{Parampreet Singh  }
\address{Perimeter Institute for
Theoretical Physics, 31 Caroline Street North, Waterloo, Ontario
N2L 2Y5, Canada}
\ead{psingh@perimeterinstitute.ca}

\begin{abstract}

A unified treatment of all known types of singularities 
for  flat, isotropic and homogeneous spacetimes in the framework of loop quantum cosmology (LQC) is presented. These include bangs, crunches and all future singularities.
Using effective spacetime description we perform a  model independent   general analysis of the 
properties of curvature, behavior of geodesics and  strength of singularities. For illustration purposes a phenomenological model based analysis is also performed.
We show that all values of 
the scale factor at which a strong singularity may occur are excluded from the effective loop quantum 
spacetime. Further, if the evolution leads to either a vanishing or divergent scale factor then the loop quantum universe is asymptotically deSitter in that regime. We also show that there 
exist a class of sudden extremal events, 
which includes a recently discussed possibility, for which the curvature or its derivatives will always diverge. 
Such events however turn out  to be harmless weak curvature singularities beyond which geodesics can be extended. Our results point towards a generic resolution of physical singularities in LQC.

\end{abstract}
\maketitle

\section{Introduction}

Singularities are common  in General Relativity (GR). These 
are the  boundaries of spacetime 
which can be reached by an observer in a finite proper time where 
 the spacetime curvature and tidal forces become infinite. 
Singularity theorems of Penrose, Hawking and Geroch show that  the primary characteristic 
of a physical singularity is the inextendibility of the geodesics beyond it \cite{penrose-hawking,geroch}. However behavior of geodesics is insufficient to capture the detailed features of singularities and distinguish physical from unphysical ones. Hence singularities are also  classified in terms of strong and weak types \cite{ellis-schmidt,tipler,krolak}. A singularity is strong  if the tidal forces 
cause complete destruction of objects irrespective of their physical characteristics, whereas a singularity is considered weak if tidal forces are not strong enough to forbid passage of objects or detectors. Due to this reason only strong singularities are generally considered as physical. An example of strong singularity is the big bang singularity in cosmological models and an example of a weak singularity is the shell crossing singularity in gravitational collapse scenarios where even 
though curvature invariants diverge, `strong detectors' can pass the extremal event \cite{seifert}.

Traditional cosmological singularities, such as big bang and big crunch,  come with 
certain signatures: events where the scale factor goes to zero, geodesics are incomplete and objects are crushed to zero volume by 
infinite gravitational curvature. However, recently various new singularities have been discovered in GR \cite{bigrip,sudden}\footnote[1]{Similar solutions have also been found in  braneworld models \cite{varun}.}. These singularities which are typically investigated in the future evolution of the universe (hence popularly known as future singularities), do not occur at a vanishing scale factor.
 The latter either goes to infinity in finite proper time, along with 
a similar behavior of energy density $\rho$ and pressure $P$ (the Big Rip) or the singularity is sudden i.e. at a finite value of time and scale factor, curvature or one of its higher derivative blow up \cite{sudden,sudden1,future}. Features of cosmological singularities can be classified using the triplet of variables $(a, \rho, P)$ and have been understood in detail \cite{visser,lazkoz}.
 For a universe with a Robertson-Walker metric and matter 
satisfying a non-dissipative cosmological equation of state: $P = P(\rho)$, cosmological singularities apart from the big bang and big crunch  
can be completely classified in four types \cite{not}: Type I as big rip, type II as the sudden one where energy density is finite but the pressure diverges at the extremal event, type III where both energy density and pressure diverge at a finite value of scale factor and rest of the remaining  as type IV where curvature components are finite but their higher derivatives blow up.

An open question is the way quantum gravitational effects change the picture 
near above extremal events. This can be asked in different ways: Does quantum gravity resolve all  spacelike singularities?
Do quantum gravity effects always bound the spacetime curvature? Are geodesics complete (if their notion exists in a quantum spacetime)? What does the singularity resolution or the lack of it tell us about the underlying theory? Since we do not yet have a 
complete theory of quantum gravity these questions cannot yet be answered in full generality.
However, they  can be posed for cosmological singularities in a simplified setting such as a 
homogeneous universe where mini-superspace quantizations are available.

Loop quantum cosmology (LQC) \cite{lqc} is one of the settings where such questions 
can be answered. It is a non-perturbative canonical quantization of homogeneous and isotropic spacetimes 
based on loop quantum gravity (LQG) \cite{lqg}. The classical phase space variables are the 
Ashtekar connection and the conjugate triad. The elementary variables used for quantization are the holonomies of the connection and fluxes of the triad. 
The quantization follows the Dirac's program for constrained systems 
where a physical Hilbert space is obtained by solving all of the constraints 
at the quantum level and predictions are extracted via Dirac observables. In LQC, due to underlying symmetries of the homogeneity and isotropy the 
only non-trivial constraint is the Hamiltonian constraint. This is expressed in terms of holonomies and fluxes and then is quantized.
 The quantization has been successfully and rigorously completed in various interesting cases which include spatially flat FRW  
models with at least one massless scalar \cite{aps,aps1,aps2}, with and without cosmological constant \cite{bp,ap}, closed \cite{polish1,apsv} and open universes \cite{kv}
as well as the inflationary spacetimes \cite{aps3}. All these models classically exhibit big bang (and in some cases, also  big crunch) 
singularity which is generically resolved in LQC. As an example, backward evolution of states which correspond to an expanding macroscopic universe at late times results in a quantum bounce to a contracting universe when energy density becomes equal to a critical value $\rcr \approx 0.41 \rho_{\mathrm{Pl}}$.
Various results have been shown to be quite robust using an exactly solvable model in LQC (sLQC) with massless scalar as the matter content \cite{slqc}. In particular, one can show that the 
bounce is a generic property of states in the physical Hilbert space 
and that $\rcr$ is the supremum of the spectrum of the energy density 
operator on the physical Hilbert space\footnote[2]{The boundedness of the density operator also holds for a wide class of lapse function \cite{polish2}.}. The universe in the branch 
preceding ours has been shown to retain semi-classicality \cite{recall} and the quantization turns out 
to be unique if one demands consistency and physical viability of the quantization scheme \cite{cs2}.

An interesting feature of loop quantum cosmology is the availability of an effective spacetime description \cite{vt,vt1}. Numerical simulations of the exact LQC equations for universes which grow macroscopic at late times confirm that 
the effective modified Friedman dynamics captures the underlying quantum dynamics and the bounce 
to an amazing accuracy \cite{aps1,aps2,apsv,bp,ap,kv}. 
These results were obtained using massless scalar field with and without cosmological constant and also the inflationary potential. Assuming that key features of the effective dynamics  can be trusted more generally, various 
 interesting results have been obtained.
These include the resolution of big crunch singularities for 
negative potentials in Cyclic models \cite{cyclic} as well as in  string inspired pre big bang scenarios \cite{pbb} and 
relation with Palatini theories \cite{palatini}.

In the dark energy scenarios, phantom field models  have been analyzed at an effective level and are shown to be 
generically free of big rip or the type I singularity \cite{phantom, phantom1}.
An interesting example which is studied in this context is the case of a 
phantom model with an unbounded negative (positive) potential for a canonical (phantom) field \cite{singlqc}. Classical dynamics for this model predicts a 
big rip singularity in future. Using effective dynamics of LQC, authors of Ref. \cite{singlqc} find
that as the future singularity is approached, though the energy density is bounded, the  pressure and the rate of change 
of the Hubble rate blow up in LQC and they conclude that the ``singularity'' is unavoidable. This case is 
interesting because it is an explicit example where the curvature invariant is not bounded by the quantum 
geometric effects. For the particular model considered in Ref. \cite{singlqc}, quantum geometric effects convert the classical type I singularity in to a sudden type II singularity. As we will show, the latter ``singularity''  is weak and unphysical.

Although the fate of singularities has been studied for specific models in LQC, a general treatment has so far been unavailable. Further, 
details about the nature and strength 
of the various singularities and the properties of geodesics in the effective spacetime were so far not investigated. Given that an effective spacetime description is available in LQC, all these issues can be addressed and analyzed in detail. This is not only important to 
distinguish physical singularities from unphysical ones but also to understand the way loop quantization affects the fate of spacetime at these singular events. 
 Moreover the examples which have so far been studied do not fully 
exhaust all the possibilities which include type III and type IV singularities. The aim of this work is to understand all these issues for all cosmological 
singularities in flat (k=0) model of LQC with matter satisfying a non-dissipative cosmological equation of state $P = P(\rho)$. Our work will assume (i) that the effective spacetime description is valid for all matter staisfying above equation of state and (ii) the effective value of geometrical and physical entities such as geodesics and curvature invariants coincide with those derived from the 
effective spacetime metric.\footnote[4]{In the models with non-vanishing intrinsic curvature, the following analysis will also require implementation of inverse scale factor effects which modify the matter conservation equation and may become dominant when scale factor approaches zero. In the flat model, unless one restricts to a compact topology, these effects are argued to be unphysical \cite{aps2} and are not considered here.}

We organize this paper as follows. In the next section we briefly revisit loop quantization and the effective Friedman 
dynamics in LQC (for  details we refer the reader to Ref. \cite{cs2}).  In Sec. III we review 
all the singularities in $k=0$ FRW model and derive the geodesic equations for the Robertson-Walker metric and state the Clarke-Kr\'{o}lak conditions \cite{clarke-krolak}
which are necessary as well sufficient for a singularity to be considered strong a la Tipler \cite{tipler} and Kr\'{o}lak \cite{krolak}. In Sec. IV we consider a model  which is sufficiently general to illustrate the singularity resolution in LQC. This model was introduced in the context of dark energy scenarios \cite{not} and 
exhibits all possible cosmological singularities including both strong and weak. We then show that in this model all  strong singularities are resolved. The weak singularities however remain unaffected. 
In Sec. V we provide a model independent  analysis 
using effective dynamics of LQC. Here we first show that in the effective spacetime of LQC, if the evolution is such that the scale factor either vanishes or becomes infinite then the universe always approaches a deSitter state.
Cosmological observers in an effective loop quantum spacetime take infinite proper time 
to reach above values of the scale factor. As in GR, these  are nonsingular. In most cases loop quantum evolution does not lead to an asymptotic deSitter regime and these are carefully analyzed  using Lipshitz conditions for dynamical as well geodesic equations. Using the fact that energy density and hence Hubble rate are always bounded above in LQC, it is shown that  
dynamical and geodesic equations never break down. For above cases geodesics can be extended to arbitrary values of the affine parameter. {\it We show that all strong singularities are generically resolved in flat and isotropic LQC.}  The only possible ``singularities'' are weak 
curvature type. These are harmless as geodesics can be extended beyond them.
We summarize the results with a discussion in Sec VI.

\section{Preliminaries}

We will consider the dynamical features of $k=0$ homogeneous and isotropic universe in LQC. The 
3+1 spacetime is described by the manifold $\Sigma \times \mathbb{R}$, where 
$\Sigma$ is the non-compact spatial manifold, and 
the Robertson-Walker metric
\be\label{metric}
d s^2 = - \, d t^2 \, + \, a^2(t)\left(d r^2 + r^2\left(d \theta^2 + \sin^2\theta d \phi^2\right)\right)
\ee
where $a(t)$ is the scale factor of the universe. 
Here we have chosen lapse $N$ to be equal to unity so that $t$ is the proper time. 

Effective description for the loop quantization of above spacetime 
can be obtained using geometric formulation of the quantum theory. Using coherent state techniques it is possible to derive an effective Hamiltonian (up to controlled higher order corrections) for various matter sources \cite{vt}. It turns out that for states which correspond to a macroscopic universe, such as ours, at late times the following effective Hamiltonian captures the underlying loop quantum dynamics:
 \be \label{effham}
{\cal H}_{\mathrm{eff}} = - \f{3}{8 \pi G \gamma^2} \, \f{\sin^2(\lambda \b)}{\lambda^2} V +
\, {\cal H}_{\mathrm{matt}} ~ ~.
\ee
Here $\beta$ and volume $V = a^3$ are conjugate variables satisfying 
\be
\{\beta,V\} = 4 \pi G \gamma 
\ee
with $\gamma \approx 0.2375$ as the Barbero-Immirzi parameter. In the classical theory the phase 
space variable $\beta = \gamma \dot a/a$. The parameter $\lambda$ captures the 
discreteness of the underlying quantum geometry and its value is determined by the minimum eigenvalue of the area operator in loop quantum gravity (LQG) \cite{abhay-wilson},
\be
\lambda = 2 (\sqrt{3} \pi \gamma)^{1/2} \lp ~.
\ee
We consider matter  to be minimally coupled and homogeneous. In particular it satisfies  
 a cosmological equation of state $P = P(\rho)$ where 
$P$ is its pressure and $\rho$ is its energy density. 

It is to be noted that  
though there may exist further state dependent quantum corrections to the effective Hamiltonian, the numeric simulations which have so far been performed show that they turn out 
to be negligible for  states representing realistic universes we are interested in (see for example Refs. \cite{apsv,bp,ap,aps3} and also refs. \cite{lukasz,madrid} for anisotropic models). Guided by these results our analysis will assume the existence 
of above effective Hamiltonian for general matter. 

Given eq.(\ref{effham}), it is  straight forward to find the modified Friedman dynamics. The vanishing of the Hamiltonian 
constraint ${\cal H}_{\mathrm{eff}} \approx 0$ leads to
\be\label{eq:sin_rho}
\f{\sin^2(\lambda \b)}{\lambda^2} = \f{8 \pi G \gamma^2}{3} \, \rho
\ee
where $\rho =  {\cal H}_{\mathrm{matt}}/V$ is the energy density. Then from the Hamilton's equation
\ba
\dot V &=&   \nonumber \{V,{\cal H}_{\mathrm{eff}}\} = - 4 \pi G \gamma \f{\partial}{\partial \beta} {\cal H}_{\mathrm{eff}} \\
&=& \f{3}{\gamma} \, \f{\sin(\lambda \beta)}{\lambda} \, \cos(\lambda \beta) \, V
\ea
we can obtain the modified Hubble rate
\be \label{fried}
H^2 = \f{\dot V^2}{9 V^2} = \f{8 \pi G}{3} \, \rho \left(1 - \f{\rho}{\rcr}\right)
\ee
where $\rcr$ is given by
\be\label{rhocrit}
\rcr  = 3/(8 \pi G \gamma^2 \lambda^2) \approx 0.41 \rho_{\mathrm{Pl}} ~.
\ee

A similar calculation for the second Hamilton's equation: $\dot \beta = \{\beta, {\cal H}_{\mathrm{eff}}\}$ results in the modified Raychaudhuri equation
\be\label{rai}
\f{\ddot a}{a} = - \f{4 \pi G}{3} \, \rho \, \left(1 - 4 \f{\rho}{\rcr} \right) - 4 \pi G \, P \, \left(1 - 2  \f{\rho}{\rcr} \right) . ~
\ee
Since quantum geometry does not affect the matter part, the Hamilton's equation for matter field 
yield the conservation law
\be\label{cl}
\dot \rho + 3 H \, (\rho + P) = 0
\ee
where pressure $P = - \partial H_{\mathrm{matt}}/\partial V$. It is straight forward to check that
Eqs. (\ref{fried}), (\ref{rai}) and (\ref{cl}) form a closed set.

The modified Friedman and Raychaudhuri equations are sufficient to determine the non-trivial components of the 
Ricci (and the Einstein tensor) on the FRW background with modified expansion rate of the scale factor. The Ricci curvature invariant turns out to be 
\be
\label{Ricci}
R = 6 \left(H^2 + \f{\ddot a}{a} \right) =  8 \pi G \rho \, \left(1 - 3 w + 2 \f{\rho}{\rcr} \left(1 + 3 w \right) \right) .
\ee
where $w$ is the equation of state of the matter component $w = P/\rho$. 

Classical equations for the FRW spacetime can be obtained 
from 
Eqs.(\ref{fried},\ref{rai}) and ({\ref{Ricci}) in the limit $\lambda \rightarrow 0$ (that is $G \hbar \rightarrow 0$): 
\be\label{cfried}
H^2 = \f{8 \pi G}{3} \, \rho ~,~~ \f{\ddot a}{a} = - \f{4 \pi G}{3} \, (\rho  + 3 P)  ~
\ee
and 
\be
\label{cRicci}
R = 8 \pi G (\rho - 3 P) ~.
\ee

The effective dynamical equations immediately lead to an upper bound on the 
energy density and hence the Hubble rate in LQC. As we will show in Sec. IV and Sec. V, these 
features play an important role in absence of physical singularities in a 
loop quantum universe.

\section{Cosmological Singularities: Nature and Strength}

Singularities for the homogeneous and isotropic spacetime can be classified using the behavior of scale factor, energy density and pressure (or equivalently in terms of spacetime curvature). All 
of the known (and plausible) singularities for matter satisfying non-dissipative equation of state $P = P(\rho)$ fall in one of the categories 
below \cite{visser,lazkoz,not}:

{{\it Big Bang and Big Crunch:}} These are accompanied by vanishing 
of the scale factor at a finite proper  time and the divergence in energy density and curvature invariants. %
Null energy condition (NEC), $(\rho + P) \geq 0$, is always satisfied in these events. \\

{\it Big Rip or Type I singularity:} For these singularities NEC and hence all other energy conditions are  violated. The scale factor diverges in proper finite time, $a(t) \rightarrow \infty$. This is accompanied with a divergence of energy density, pressure and 
curvature invariants. \\

{\it Sudden or Type II singularity:} This extremal event occurs at a finite 
value of the scale factor $a \rightarrow a_e$. It is characterized by a finite value of the 
energy density but an associated divergence of pressure. Due to the latter,  
$R$ diverges. \\

{\it Type III singularity:} As type II singularity, this singularity 
also occurs at a finite value of the scale factor. However, both the energy density and pressure diverge, causing a blow up of curvature invariants. \\

{\it Type IV singularity:} None of the energy density or pressure blow up in this case which occurs at a finite value of the scale factor. Curvature invariants are finite, however
curvature derivatives blow up. 

Singularities associated with a divergence in spacetime curvature fall in either big bang/crunch or type I-III class. 
Remaining singularities are curvature derivative kind which form Type IV class. 
Unlike big bang/crunch and 
type I singularities, 
analysis of energy conditions for type II, III and IV is subtle and answers 
can be 
model dependent \cite{visser}. However, it is always true that 
Type II singularities are accompanied by violation of dominant energy condition (DEC): 
$(\rho \pm P) \geq 0$. 

To understand the behavior of geodesics let us consider the geodesic equations for the flat $(k=0)$ Robertson-Walker metric:
\be
(u^{\alpha})^\prime + \Gamma^\alpha_{~ \beta \nu} \, u^\beta u^\nu = 0 ~
\ee
where prime denotes a derivative 
with respect to the affine parameter $(\tau)$.
Using  Cartesian coordinates, $u^x, u^y$ and $u^z$ turn out to be constant. 
Therefore it is sufficient to analyze geodesic equation for  
time coordinate which is
\be\label{g1}
t^{\prime \, 2} = \epsilon + \f{\chi^2}{a^2(t)} 
\ee
where $\chi$ is a constant and $\epsilon = 1$ for massive particles and $\epsilon = 0$ for null geodesics. Since below we also consider radial geodesics, it is also useful to obtain equation for radial 
geodesics
\be\label{g0} 
a^2(t)\,\, r^\prime = \chi
\ee
which using (\ref{g1}) implies 
\be\label{g2}
\ppt = - a \dot a \, r^{\prime \, 2} = - H \, (t^{\prime \, 2} - \epsilon) ~.
\ee

Understanding properties of above geodesic equations  is important to prove whether 
the spacetime is geodesically complete. For that it is necessary to show that 
geodesic equations admit a unique extendible solution.
From the geodesic equations we see that these break down when scale factor 
becomes zero and/or the Hubble rate blows up. Hence these break down at big bang/crunch, type I and 
type III singularities. Since type II and type IV singularities occur at a finite value of scale factor and Hubble rate remains finite at these events, therefore 
geodesic equations do not break down. We will later show that for these cases 
a unique extendible solution to the 
past and future of type II and type IV singularities can be found (see also Ref. \cite{lazkoz2}).
 Thus, type II and type IV singular events fail to be singularities a la theorems of Penrose, Hawking and Geroch. From the criteria of geodesic inextendibility only 
big bang/crunch, type I and type III singular events turn out to be singularities. 

Apart from analysis of geodesics it is also useful to consider the strength of the singularities. A detailed discussion of the strength of various cosmological singularities is available in Ref. \cite{lazkoz}. We here only note the conditions necessary for  
our analysis. 
These originate from the work of Clarke and Kr\'{o}lak to differentiate types of singularities which involve analysis of integrals of curvature components for both null and particle geodesics \cite{clarke-krolak}.
For the FRW metric,  a singularity occurring at the value of the affine parameter $\tau = \tau_e$ is a strong curvature  type  a la Tipler iff the following 
integral over the spatial components of Ricci tensor is unbounded 
\be\label{tipler}
\int^\tau_0 d \tau' \int^{\tau'}_0 \, d \tau'' R_{ab} u^a u^b ~,
\ee
as $\tau \rightarrow \tau_e$. Else  the singularity is weak. A less restrictive condition is by Kr\'{o}lak who classifies the singularity to be strong iff
\be\label{krolak}
\int^\tau_0 d \tau R_{ab} u^a u^b ~
\ee
is infinite as $\tau \rightarrow \tau_e$. It is thus possible that 
a singularity may be strong a la Kr\'{o}lak but weak a la Tipler. 
A strong singularity from above conditions is the one in which an in-falling 
observer or detector is completely annihilated by the tidal forces. For a 
weak curvature singularity, tidal forces are not strong enough to cause 
such a destruction. Sufficiently strong detectors survive such events.

Note that the integrand in both of above integrals is  proportional to combination of square of Hubble rate and 
$\ddot a/a$. As an example, for null geodesics
\be
R_{ab} \, u^a u^b = 2 \f{\chi^2}{a^2(t)} \left(H^2 - \f{a(t)''}{a(t)}\right)
\ee
where we have used (\ref{g1}) and (\ref{g0}).
Since conditions (\ref{tipler}) and (\ref{krolak}) involve at least one  integral over affine parameter, it turns out that the integrals are finite  if only  
$\ddot a$ or higher derivatives diverge and the scale factor neither vanishes nor diverges. This happens in the case of type II and type IV singularities which are thus weak.  
Big bang, big crunch and big rip are strong curvature singularities  according to both Tipler and Kr\'{o}lak. Type III singularities are strong according to Kr\'{o}lak's condition but weak by Tipler's condition \cite{lazkoz}. (Similar conclusions are reached by the analysis of particle geodesics).
It is to be noted that strong curvature singularities in FRW universe are also the ones beyond which geodesics can not be extended. On the other hand weak singularities are the ones beyond which geodesics can be extended and thus are harmless events.

\section{Singularity resolution in LQC: Illustration via a Model}
We now illustrate loop quantum dynamics and results of the previous section 
using a general model which exhibits all the cosmological singularities of interest \cite{not}. This model is of interest because it describes a general dark energy scenario in a FRW universe including quintessence and phantom dark energy models. Classically the model predicts various singularities for different ranges of parameters and it proves useful to understand 
the detailed properties and the fate of the universe in such scenarios.
The model is based on the ansatz 
\be\label{not_pressure}
P = - \rho  - f(\rho)
\ee
with
\be
f(\rho) = \f{A B \rho^{2 \alpha - 1}}{A \rho^{\alpha - 1} + B} ~.
\ee
Here $A$, $B$ and $\alpha$ are  parameters of the model. Their values 
determine the nature of singularities. 
Note that when $f(\rho) = 0$, the model reduces to the standard cosmological constant scenario. Above anasatz therefore proves very useful to study departures of the equation of state from the cosmological constant setting.

The dependence of energy density on scale factor can be found by 
integrating (\ref{cl})
\be
a = a_o \, \exp\left(\f{1}{6} \f{(2 A + B \rho^{(1 - \alpha)}) \rho^{1 - \alpha}}{A B (1 - \alpha)}\right)
\ee
which yields
\be
\rho = \left(-\f{A}{B} \pm \left(\f{A^2}{B^2} - 6 (\alpha - 1) A \ln \left(\f{a}{a_o}\right)\right)^{1/2}\right)^{1/(1 - \alpha)} ~.
\ee

To investigate  resolution of various singularities we will use 
eqs.(\ref{fried}), (\ref{rai}), (\ref{Ricci}) along with $\dot R = 6(\ddot H + 
4 H \dot H)$ and 
\be
\dot P =  \pm \sqrt{24 \pi G \rho (1 - \rho/\rcr)} (\rho + P) \, \bigg[1 ~ + ~ \f{(2 \alpha - 1) \, A B \, \rho^{2 \alpha - 2}}{A \rho^{\alpha - 1} + B} + ~\f{(1-\alpha) A^2 B \rho^{3 \alpha - 3}}{(A \rho^{\alpha - 1} + B)^2} \bigg] ~.
\ee
In the following we will only consider 
the future evolution and compare the results of the classical and loop quantum evolution. (For details of the classical dynamics for various choices of  parameters of this model we refer the reader to Ref. \cite{not}.).

\begin{figure}[]
\includegraphics[scale=0.7]{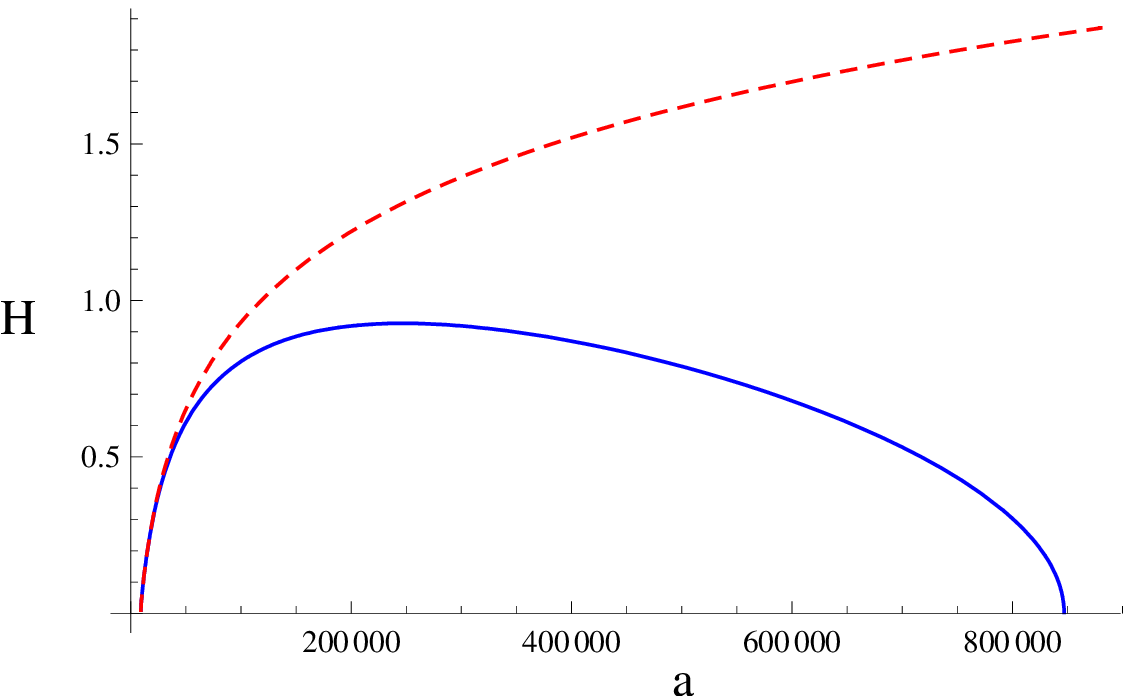}
\includegraphics[scale=0.7]{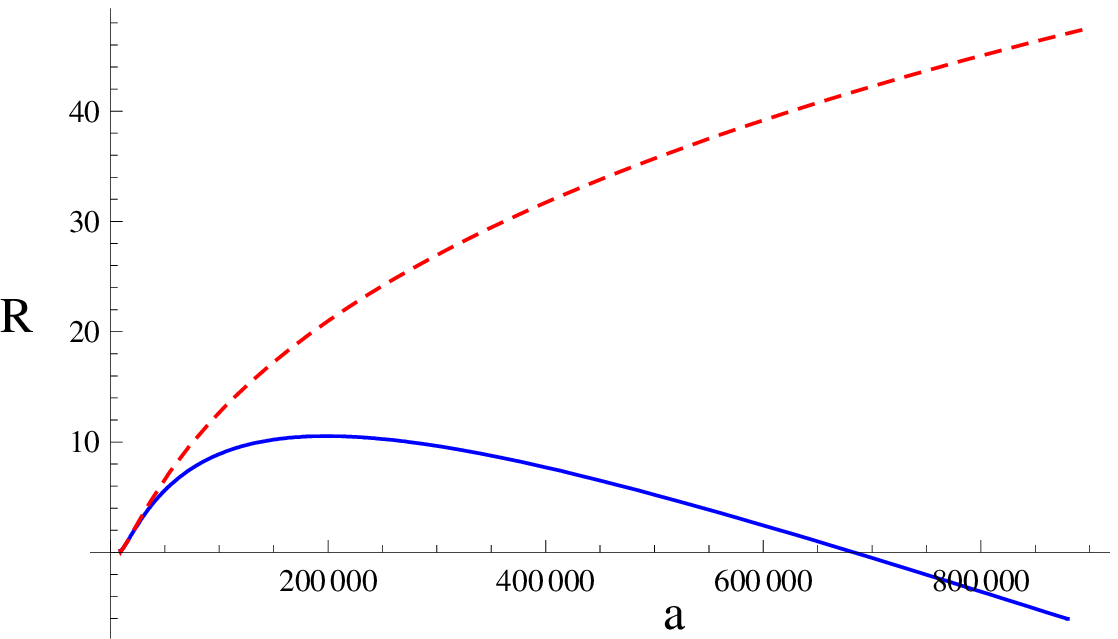}
\caption{Classical (dashed) and effective LQC (solid) curves of Hubble rate and  Ricci scalar for type I singularities. The values of parameters are $A = 0.1, B = -1$ and $\alpha = 0.8$. }
\label{t1_1}
\end{figure}

\subsection{Type I Singularities}
If the value of $\alpha$ is chosen between $3/4 < \alpha < 1$ and $A$ is 
positive, then the model gives a big rip (type I) singularity in  GR. The scale factor, energy density and pressure diverge  at a finite time and the DEC is violated (for all times). There is no big bang in the classical theory 
(since DEC is violated). The model is devoid of an initial 
singularity.

In LQC, the big rip singularity is avoided. The energy density initially grows 
as in the classical theory, however when it becomes comparable to $\rcr$, departures from classical trajectories become significant. Eventually, $\rho$ 
becomes equal to $\rcr$ and the Hubble rate vanishes with $\ddot a$ taking negative value.
The universe instead of ripping apart 
in finite time, recollapses and the evolution continues. The Ricci scalar, its 
derivatives and higher curvature invariants are bounded in the entire evolution.

In Fig. \ref{t1_1} we compare the evolution of Hubble rate and Ricci curvature scalar in the classical 
and the effective dynamics of LQC. 
The classical Hubble rate diverges 
as $a \rightarrow \infty$, whereas the Hubble rate in LQC is bounded. 
As can be seen, unlike in GR, $R$ is bounded in the loop cosmological evolution.

\subsection{Type II Singularities}
A necessary condition for type II singularities to occur is $A/B < 0$. For 
these cases there is also a big bang (crunch) singularity as $a \rightarrow 0$.
The sudden singularity occurs at a finite value of the scale factor, $a \rightarrow a_o$. As  the singularity is approached the energy density goes to zero however 
pressure and hence the Ricci curvature diverge. 
Since the Hubble rate is bounded, the geodesics are extendible and the singularity is only a weak curvature singularity.

In loop quantum evolution the initial singularity which is a strong curvature type is resolved. Since $\rho \ll \rcr$ when the sudden singularity occurs, the latter is not resolved. In fact near this extremal event the dynamics mimics 
the one obtained from  GR and the properties of geodesics do not change qualitatively near the sudden singularity. As in the classical theory, $\ddot a/a$ and $R$ 
diverge near the type II event. Since the initial big bang singularity is also resolved  the effective spacetime in LQC is geodesically complete.

Dynamical trajectories obtained from numerical integration of classical and loop quantum equations are compared in Fig. \ref{t2_1}. 
The classical Hubble rate  (dashed) diverges at the big bang 
and approaches zero near $a_o$. The loop quantum Hubble rate (solid curve) is bounded throughout the evolution and agrees with the classical values near $a \rightarrow a_o$. It vanishes at the bounce point in the early universe and 
approaches the classical value at late times. The behavior of Ricci scalar shows that it is bounded in LQC
in the early epoch and attains a positive value at the quantum bounce. However, it 
diverges when the sudden singularity is approached. The quantum geometric effects are unable to bind the curvature  in this case.

\begin{figure}[tbh!]
\includegraphics[scale=0.7]{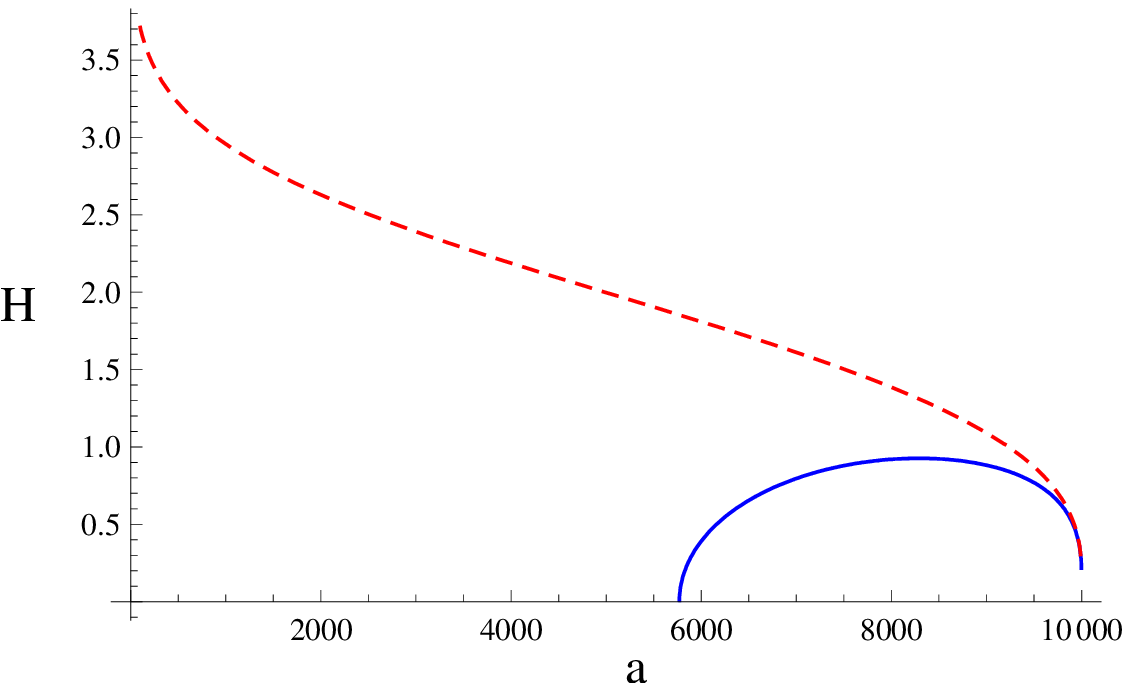}
\includegraphics[scale=0.7]{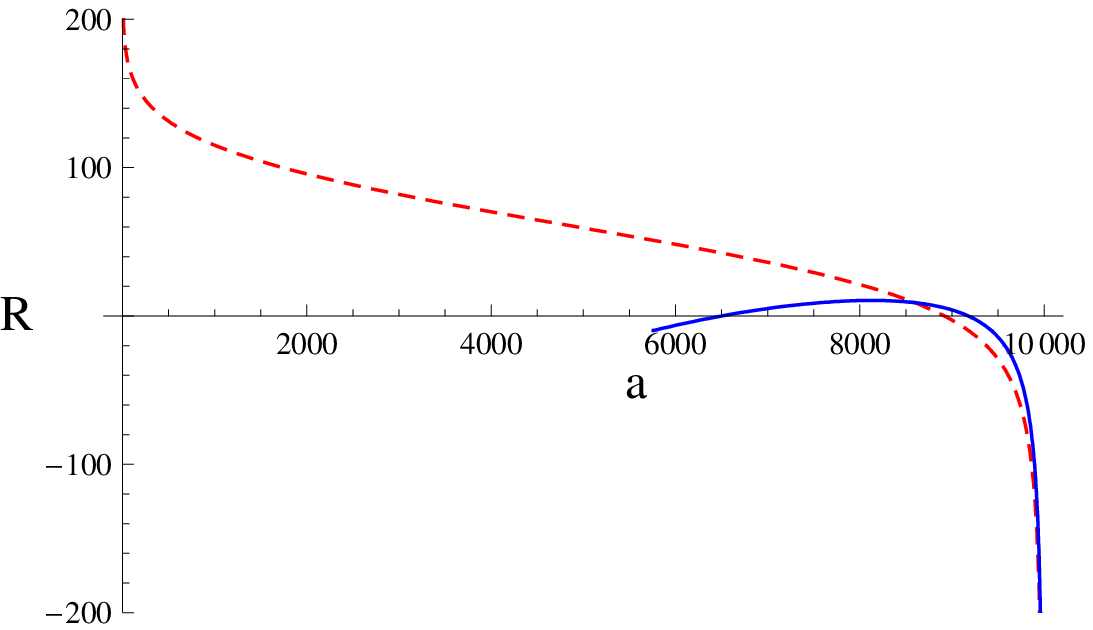}
\caption{Plot of Hubble rate and Ricci scalar for type II singularity. The classical Hubble rate diverges at the big bang but is finite (goes to zero) at $a = a_o$. 
Parameters are $A = -0.1, B = 1$ and $\alpha = 1/4.1$. }\label{t2_1}
\end{figure}

Since pressure grows unboundedly as the type II singularity is approached, there is a huge violation of DEC near $a_o$. For $\alpha < 0$, using (\ref{not_pressure}) we obtain: $w \rightarrow - \infty$ for positive B when $a \rightarrow a_o$.
It can also be shown  that these extremal events do not satisfy 
Tipler and Kr\'{o}lak's conditions for strong singularities. For null geodesics  we obtain 
\be
R_{ij} u^i u^j = 8 \pi G (\rho + P) \f{\chi^2}{a^2} \left(1 - 2 \f{\rho}{\rcr}\right) ~,
\ee
using which we can compute the Tipler \cite{tipler} and Kr\'{o}lak \cite{krolak} integrals. Both of these are finite in LQC. As an example we show the 
integrand of (\ref{krolak}) (after a change of variables to the scale factor) in 
Fig. \ref{integral}. Value obtained from numerical integration for the chosen values of parameters turns out to be approximately $5.6 \times 10^{-4}$. Similar results follow  for the particle geodesics.

\begin{figure}[]
\hskip4cm \includegraphics[scale=0.7]{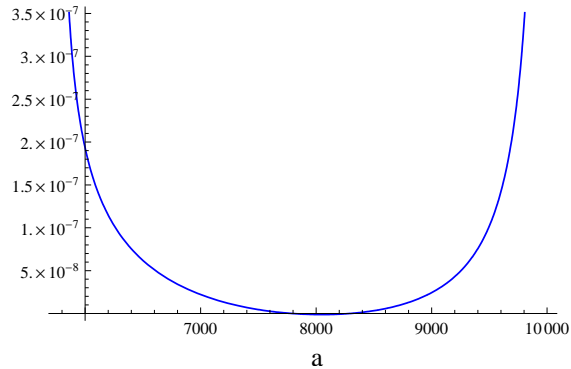}
\caption{The integrand of (\ref{krolak}) for the values of parameters in 
Fig. \ref{t2_1} is shown for effective LQC. The numerical integration gives 
a finite answer.}
\label{integral}
\end{figure}
\begin{figure}[tbh!]
\includegraphics[scale=0.7]{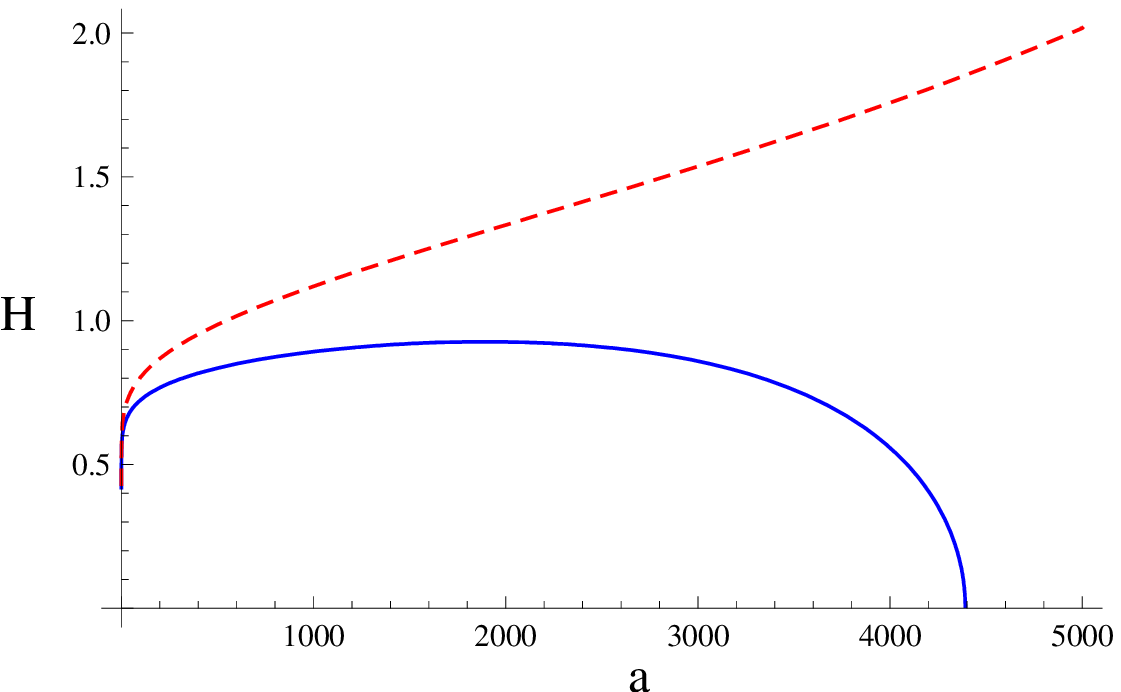}
\includegraphics[scale=0.7]{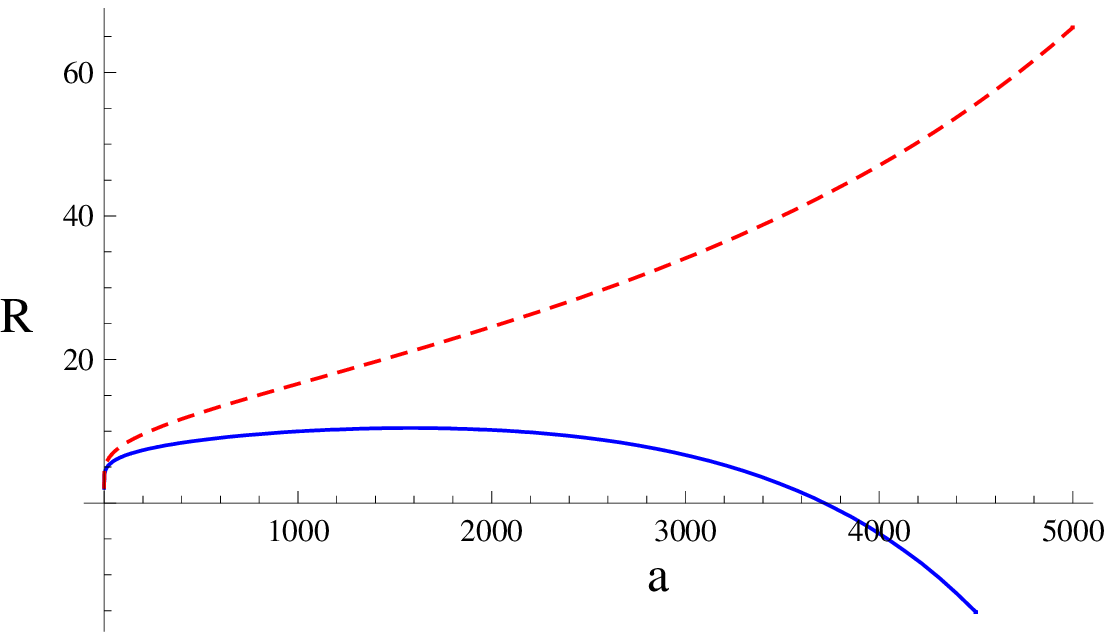}
\caption{The classical (dashed) Hubble rate and Ricci scalar is compared with the one in LQC for 
type III extremal events. Classically Hubble rate blows up as the singularity 
is approached at late times. Both the Hubble rate and Ricci are bounded in LQC. Parameters are $A = 100, B = 1$ and $\alpha = 2$.}
\label{t3_1}
\end{figure}

\subsection{Type III Singularities}
When $\alpha > 1$, the model predicts a type III singularity at a finite time 
and scale factor $a \rightarrow a_o$. Both the energy density and pressure 
diverge in GR. The unbounded Hubble rate causes geodesics to be incomplete and 
results in a strong curvature singularity a la Krolak. Since the singularity 
occurs at a finite volume, it is weak according to Tipler's criteria. 
In this model the initial singularity is absent at the classical level itself.

In loop quantum dynamics there is no type III singularity. When $\rho$ is small, the classical and effective evolutions are similar. However they become qualitatively different as $\rho$ increases. The loop quantum universe recollapses at 
$\rho = \rcr$ and  
the type III singularity is avoided. The Hubble rate, $\ddot a/a$ and Ricci scalar are bounded and finite. Thus geodesics are complete and integrals (\ref{tipler}) and (\ref{krolak}) are finite.

A comparison of the classical and effective LQC evolution is presented in Fig. \ref{t3_1}. 
 As can be seen in the plot
the Hubble rate is bounded in LQC, whereas it grows unbounded classically in the future evolution. Similarly, the Ricci scalar which diverges in the classical theory as $a_o$ is 
approached is bounded in LQC. \\

\subsection{Type IV Singularities}
If the value of $\alpha$ is between $0 < \alpha < 1/2$, then as $a \rightarrow a_o$ though the energy density and pressure remain finite, a higher derivative of 
curvature diverges. This singularity is a derivative curvature singularity and 
none of the curvature invariants diverge. As in type II case, since Hubble rate  
is finite at $a = a_o$, geodesic equations are well behaved. The singularity 
is a weak curvature type under the classifications of both Tipler and Kr\'{o}lak.

\begin{figure}[tbh!]
\includegraphics[scale=0.7]{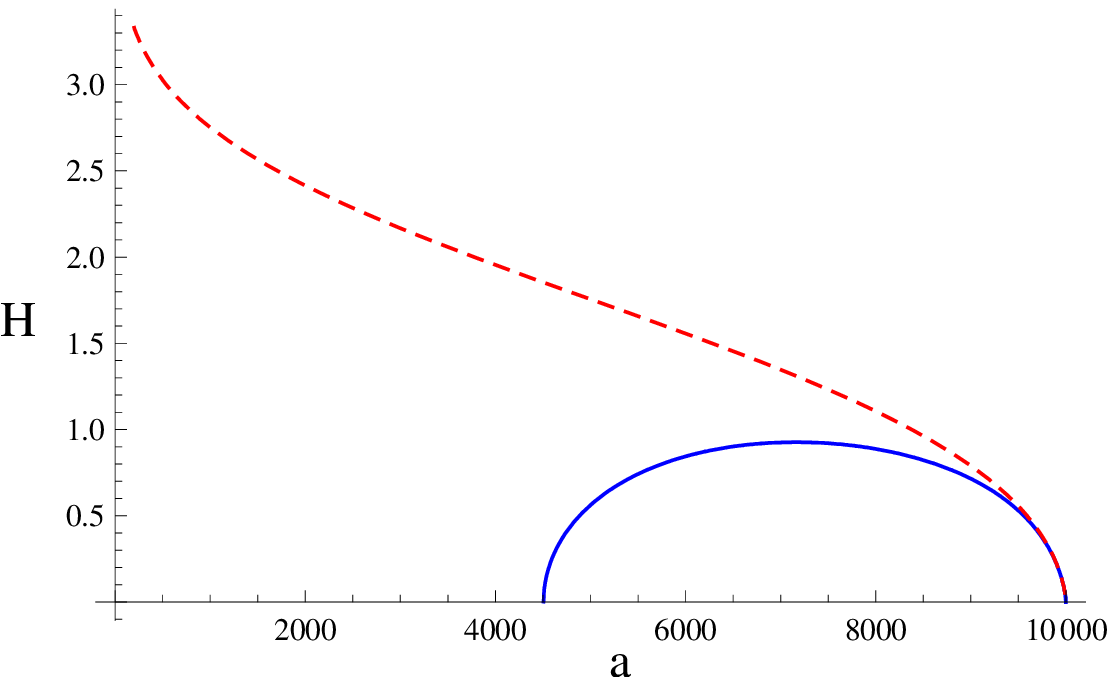}
\includegraphics[scale=0.7]{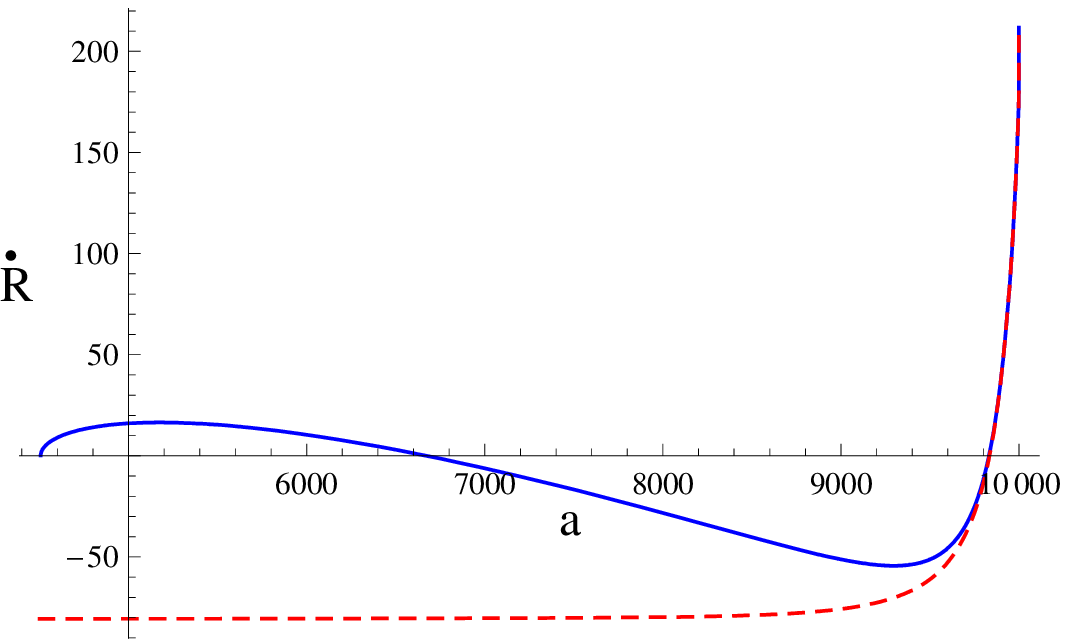}
\caption{Comparison of Hubble rates and $\dot R$ is made for the classical theory and LQC.
The classical divergence is controlled by the quantum geometry leading to a 
bounce in the early universe. At late times, the classical curve is a good 
approximation to LQC and both vanish at $a = a_o$. Parameters are $A = -0.1, B =-1$ and $\alpha = 1/4.1$ }\label{t4_1}.
\end{figure}

Quantum geometric effects have little influence on this harmless extremal event beyond which geodesics can be extended 
even in the classical theory.
However they do resolve the big bang singularity for this model which accompanies a classical divergence of Hubble rate and Ricci curvature and thus lead to a geodesically complete spacetime.
As is evident from the Fig. \ref{t4_1}, Hubble rate is finite 
through out the loop quantum evolution.

For type IV singularities, the value of $\alpha$ determines the order of 
derivative which blows up at $a = a_o$ \cite{not}. For $\alpha = 1/4.1$, the divergence occurs in $\ddot H$ and hence $\dot R$ which is depicted in the second plot of Fig \ref{t4_1}.

\section{Singularity resolution in LQC: General analysis and some observations}

In the previous section we saw that the effective dynamics of LQC successfully 
resolved all types of strong singularities in FRW cosmology for quite a general 
equation of state in a model proposed in Ref. \cite{not}. 
An important question is whether these results hold in general for a cosmological equation of state of the form $P = P(\rho)$. 
To understand this let us first note that modified Friedman dynamics 
in LQC (\ref{fried}) leads to a universal upper bound $\rcr$, i.e.
\be
0 \,  < \, \rho \, \leq \, \rcr ~
\ee
and the Hubble rate is bounded with the maximum allowed value at $\rho = \rcr/2$:
\be
|H|_{\mathrm{max}} = \left(\f{1}{\sqrt{3} \, 16 \pi G \hbar \gamma^3}\right)^{1/2} ~
\ee
(where we have used the expression for $\rcr$ (\ref{rhocrit})). This implies a bound on the trace of the extrinsic curvature $K$, since  for the flat Robertson-Walker metric it is related to Hubble rate as $K = 3 H$. Note that the upper bound arises purely because of quantum gravity. If $G \hbar \rightarrow 0$ we find that $|H|_{\mathrm{max}} \rightarrow \infty$.

Since energy density is bounded above by $\rcr$, all values of the scale factor 
for which $\rho > \rcr$ are excluded from the effective spacetime of LQC. 
This immediately implies that singularities associated with divergence of energy density and Hubble rate are absent in LQC. These include big bang/crunch, type I and type III singularities i.e. all strong singularities in FRW cosmology.

Before we analyze the nature of geodesics in LQC, let us note an interesting property of the effective equations related to two values of the scale factor: $a(t) = 0$ and $a(t) = \infty$. In classical cosmology depending on the equation of state of matter, a big bang/crunch singularity may occur at $a(t) = 0$ and a big rip (type I) singularity may occur at $a(t) = \infty$.
However, in effective loop quantum spacetime this is not the case.
In LQC, if evolution leads to above values of scale factor  then the universe always approaches a deSitter state, i.e. equation of state becomes that of the positive cosmological constant ($w = -1$). From the 
modified Friedman dynamics we find that unlike classical theory, energy density, Hubble rate and curvature invariants are always finite at $a(t) = 0$ and $a(t) = \infty$ in LQC. It can also be shown that cosmological observers take infinite proper time to reach above values of scale factor in the effective loop quantum spacetime.

We prove this by contradiction. Let us first note that the following identity obtained from the conservation law (\ref{cl}) holds for all values of the scale factor: 
\be\label{intcons}
\ln \left(\f{\rho}{\rho_o}\right) = - 3 \, \int_{a_o}^{a} \, (1 + w(\tilde a)) \f{d \tilde a}{\tilde a} ~.
\ee
 We assume that $w(a)$ be a smooth real function of the scale factor. Let us now assume that it is possible to approach $a(t) = 0$ without $w \rightarrow -1$ and satisfy above equation. Since 
left hand side of above equation is finite, it requires the integrand to be such that the right hand side is also finite. This implies that the product of above integrand with scale factor should go to zero as $a \rightarrow 0$. Which means that 
$(1 + w(a))$ must vanish  in the above limit for Eq.(\ref{intcons}) to be satisfied, i.e.  $w \rightarrow -1$ as $a(t) \rightarrow 0$. If $w \neq -1$ as $a(t) \rightarrow 0$, the right hand side blows up and the equation is not satisfied. Hence our assumption is incorrect.



Similarly we can prove that in LQC it is not possible to approach $a=\infty$ without $w(a) \rightarrow -1$. To prove it by contradiction, let us assume that we can approach $a(t) = \infty$ without $ w \rightarrow -1$ and satisfy eq.(\ref{intcons}). Since the R.H.S of  (\ref{intcons}) is finite,  the L.H.S is finite only if the product of the scale factor and the integrand goes to zero as $a \rightarrow \infty$.
That is, $(1 + w(a)) \rightarrow 0$  as $a(t) \rightarrow \infty$. For equation of state not approaching $-1$ as $a\rightarrow \infty$, above equation can not be satisfied and therefore our assumption turns out to be incorrect. Hence we can state,


\begin{rem1}
In  flat isotropic LQC if the evolution leads to either a vanishing or a divergent value of 
the scale factor then the universe is asymptotically deSitter in that regime.
\end{rem1}

This  feature of LQC is very interesting. To understand it futher let us consider loop quantum dynamics with a 
positive cosmological constant $(\Lambda)$. Here we first note that for a matter as pure cosmological constant   the modified 
Friedman and Raychaudhuri equations  in LQC are equivalent to the  classical ones with a ``renormalized'' cosmological constant. To see this let us consider the modified Friedman and Raychaudhuri equations
for a deSitter universe sourced with energy density $\rho = \Lambda/(8 \pi G)$, 
\be
H^2 = \f{\Lambda'}{3}~~~~~~ \mathrm{and} ~~~~~~ \f{\ddot a}{a} = \f{\Lambda'}{3}
\ee
where $\Lambda'$ is the ``renormalized'' cosmological constant\footnote[6]{The ``renormalization'' of cosmological constant due to quantum geometry effects has an interesting feature. If $\Lambda \ll \sqrt{3}/(32 \pi^2 \gamma^2 \lp^2$ then the correction to cosmological constant is 
very small. However, if $\Lambda$ is of the order Planck or more precisely $\Lambda \approx \sqrt{3}/(32 \pi^2 \gamma^2 \lp^2)$ then the ``renormalized'' value of cosmological constant is very small compared to Planck scale.}: 
\be
\Lambda' = \Lambda\left(1 - \f{\Lambda}{8 \pi G \rcr}\right) ~.
\ee
 Solving above field equations we can obtain the solution to the scale factor in the small neighbourhood of $a(t) = 0$ or $a(t) = \infty$. The solution behaves as 
  $a(t) \approx \exp(\pm \sqrt{\Lambda'/3} ~ t)$ as we approach above values of the scale factor.  Hence   cosmological observers in LQC take infinite proper time to reach 
$a(t) = 0$ or $a(t) =  \infty$. As in classical GR, in these cases the spacetime is extendible (and in this sense it is non-singular).

Above cases where the evolution leads to a vanishing or a divergent scale factor are not so common in a loop quantum evolution. In most cases of interest 
evolution does not lead to above values and we now focus on them. To understand the behavior of 
 geodesics and their extendibility in these cases, it is useful to first consider the dynamical equations (\ref{fried}) and (\ref{cl}) 
and analyze  Lipshitz conditions for existence of a unique solution. It is straightforward to find that 
except the following critical points, the Lipshitz conditions are always satisfied and equations are regular. These points are (i) when energy density becomes equal to the critical energy density $\rcr$, (ii) when pressure $P$ diverges with a finite value for energy density and (iii) when $\dot P$ diverges at a finite value of energy density and pressure. First critical point corresponds to the bounce/recollapse point in an LQC universe. From the classification in Sec. III we see that the second and the third critical points correspond to type II and type IV singularities respectively.  It is important to note that none of the critical points correspond to a strong singularity. Further above critical points are not problematic. In the neighbourhood of these points we can use
\be
a = a_o \, \exp\left(-\f{1}{3} \int_{\rho_o}^{\rho} \f{d \tilde \rho}{\tilde \rho + P(\tilde \rho)}\right)
\ee
and 
\be
t = \mp \int_{\rho_o}^{\rho} \f{d \tilde \rho}{\sqrt{24 \pi G} \tilde \rho^{1/2} (1 - \tilde \rho/\rcr)^{1/2} (\tilde \rho +  P(\tilde\rho))} ~,
\ee
to determine the scale factor and obtain a unique solution $a(t)$  in the past and future of the critical points. Since Lipshitz conditions are satisfied everywhere else except above points it is hence possible to obtain a global solution for the dynamical equations. 

Now we analyze geodesic equations. 
 The upper bound on 
the Hubble rate ensures that these never break down in LQC.
Lipshitz conditions are satisfied for null and particle geodesics and hence a unique extendible solution exists. As an example, let us consider equations for null geodesics for time (\ref{g1}) and radial coordinates (\ref{g0}): $t' = \chi/a(t) = f(t,\tau)$ and $r' = \chi/a^2(t) = g(r,\tau)$. For cases under consideration: $f, g < \infty$. The 
derivative of $f$ and $g$ with respect to radial coordinate is trivially zero. 
Due to boundedness of Hubble rate,  the derivatives with respect to time are finite and thus  Lipshitz conditions are satisfied. A similar analysis can be performed for the particle geodesics and we obtain the same result.
Therefore for cases under consideration geodesics can be extended to arbitrary values of the affine parameter. 

These results should be contrasted with those in the classical theory where geodesics can be extended 
only beyond type II and type IV singularities, i.e. weak singularities \cite{lazkoz,lazkoz2}. Since geodesics can not be extended beyond strong singularities which are common in the classical theory, classical spacetimes are in general geodesically incomplete.

Let us now consider the behavior of curvature invariants. Due to underlying symmetries of the Robertson-Walker metric, it is sufficient to analyze Ricci curvature scalar $(R)$. As we have seen Hubble rate is always bounded in LQC. If ~ $\ddot a /a$ \, is also bounded, then $R$  is bounded.
 From Eq.(\ref{Ricci}) it is clear that divergence in $R$ can only arise if 
pressure and hence the equation of state diverges. This corresponds to the 
type II singularity in LQC (which was also observed in ref. \cite{singlqc}). 
Since geodesics can be extended beyond them, these are harmless events. 
Note that for any reasonable form of matter, the equation of state is expected to be finite and hence for such reasonable forms of matter curvature invariants never diverge in LQC.

Finally let us turn to the strength of these extremal events. From discussion of  Tipler (\ref{tipler}) and Kr\'{o}lak (\ref{krolak}) conditions in Sec. III we know that existence of a  strong curvature singularity requires a divergence in Hubble and/or a vanishing of the scale factor. In LQC Hubble rate is always bounded. The scale factor vanishes or diverges only when universe approaches a deSitter phase for which the integrals (\ref{tipler}) and (\ref{krolak}) are finite. Thus Tipler and Kr\'{o}lak integrals are always finite in LQC.
Hence we conclude,

\begin{rem1}
No strong curvature singularities exist in the effective spacetime of flat isotropic LQC.
\end{rem1}

It is interesting to note that loop quantum modifications only resolve 
strong singularities in cosmology whereas weak singularities may still occur in 
the effective spacetime. However as we have shown geodesics can be extended beyond the latter events and sufficiently strong detectors survive tidal forces.  Thus quantum geometry is able to 
distinguish between physically relevant strong singularities from unphysical weak ones and resolves only the former.

\section{Conclusions}
One of the key predictions of LQC is the resolution of big bang (big crunch) singularity and existence of quantum bounce  at 
Planck scale in various models \cite{aps2,bp,ap,apsv,aps3,slqc}. The underlying 
quantum dynamics for states which lead to a macroscopic universe at late times can be described by  modified Friedman 
and Raychaudhuri equations derived from an effective Hamiltonian \cite{vt}. 
The latter turn out to be very successful in capturing  details of the 
true quantum evolution, a feature which has been tested for various forms of matter ranging from equation of state 
of massless scalar to that of the cosmological constant. Modified Friedman dynamics is hence a valuable 
tool to probe the way quantum geometric effects resolve the singularities in 
the effective spacetime. An important question is whether the results of singularity resolution are generic. 
The aim of this work was to understand this issue in the effective spacetime description of flat homogeneous isotropic LQC with a general non-dissipative cosmological equation of state.

 Assuming that the effective equations are valid 
for a general matter model we analyzed in detail nature and strength of all 
possible cosmological singularities and behavior of geodesics in the $k=0$ isotropic and homogeneous 
model \footnote[8]{Some of these singularities 
may be already restricted in the full quantum theory due to properties of matter Hamiltonian. However to keep the analysis general we allowed all possibilities.}. Effective equations were used to analyze a 
 model with a general enough ansatz for equation of state which allows study of all  possible singularities in Sec. IV. A general analysis for the cosmological equation of state $P = P(\rho)$ using effective equations was performed in Sec. V. 
We found that singularities which involve divergence of energy density (or Hubble rate)
are resolved. The underlying reason is the existence of upper bound on the 
energy density of the matter  in LQC which translates in to an upper bound for the trace of extrinsic curvature. 
We show that it  results in a generic resolution 
of all strong curvature singularities and finiteness of spacetime curvature for flat isotropic LQC. Points in the classical spacetime 
where such singularities occur are  excluded in the effective spacetime of LQC.
However, there do  exist extreme events such as type II singularities  where 
the curvature diverges due to unbounded pressure  but a finite energy density. 
We showed that for such cases quantum geometry plays little role and 
divergence in curvature will not be controlled. Similarly, type IV events which involve a divergence in curvature derivative can occur in LQC. Quantum geometry does not exclude these points from the effective spacetime. Interestingly these events 
are not real singularities even in the classical theory and geodesics can be extended beyond them \cite{lazkoz,lazkoz2}.
Further, tidal forces are unable to destroy sufficiently strong detectors and 
hence these singularities are weak. 


We also found some nice properties of the effective spacetime in this analysis. First that if evolution leads to either a vanishing or a divergent values of the scale factor then the loop quantum cosmos behaves as a deSitter universe in that regime. As in the classical cosmology, the spacetime can be extended in these cases. 
Secondly, quantum geometry is 
able to distinguish physically relevant strong singularities from weak singularities. Thus, LQC 
resolves only the real singularities and ignores the harmless extremal events. In general loop quantum effects may either completely eliminate all strong curvature singularities  (as demonstrated by the model in Sec. IV) or convert them to harmless weak ones (as in the analysis of Ref. \cite{singlqc}). 
 These results are the examples that there can exist
physically interesting scenarios where divergences in curvature may not be 
regulated and yet there may be no physical singularity.

Results obtained in this work can be extended in a straightforward way to the curved spatial manifolds ($k=1$ models) and Bianchi-I anisotropic spacetimes in LQC \cite{cs3}.  It will be interesting to extend the present investigations to models which go beyond homogeneous spacetimes and also matter models with a more general equation of state than $P=P(\rho)$. Also, incorporation of higher order state dependent quantum corrections to the effective Hamiltonian is a useful direction to explore as they will give useful insights on the role of 
states in singularity resolution \cite{vt1}.
If we look at the way loop quantization is performed, the 
 deeper reason behind singularity resolution in flat cosmological model is 
 tied to the careful quantization of the Hamiltonian constraint in LQC \cite{aps2} which turns out to be unique 
in various ways \cite{cs2}.
Thus leading to a harmonious convergence of various results. It is straightforward to see that a different choice of 
quantization which is inequivalent to the improved dynamics \cite{aps2} (or sLQC \cite{slqc}) would not lead to 
generic resolution of singularity at the effective level. Hence investigations carried here have potential lessons also for the full theory in 
relation to restrictions on various quantization ambiguities.

We will also like to point out that the bound on the Hubble rate in LQC can be viewed as the one 
on the expansion of congruences in the 
effective spacetime \cite{corichi}. Whether or not this is a generic feature 
of a quantum theory of geometry is an important open question. Its answer should 
provide insights on much sought quantum generalization of the classical Raychaudhuri equation (\ref{rai}) which is crucial to prove a  non-singularity theorem \cite{naresh}.
It also remains to be seen whether the simplistic way of quantum geometry to resolve the singularities, i.e. ensuring geodesic completeness and not a bound on spacetime curvature in general and removing only strong curvature singularities, is a common feature of loop quantum spacetimes.  In our opinion this should serve as one of the guiding 
principles for such quantizations.


\ack
\noindent
We are grateful to Abhay Ashtekar, Alejandro Corichi and Tomasz Pawlowski for 
useful discussions, suggestions and comments on this manuscript. We also 
thank David Wands for comments. We further thank an anonymous referee for comments which led to improvement of this manuscript.
Research at Perimeter Institute is supported by the Government
of Canada through Industry Canada and by the Province of Ontario through
the Ministry of Research \& Innovation.

\end{document}